# Sustainability Analysis Framework for On-Demand Public Transit Systems

Nael Alsaleh[1] and Bilal Farooq[1,*]

[1]Laboratory of Innovations in Transportation (LiTrans), Department of Civil Engineering, Toronto Metropolitan University, Toronto, ON M5B 2K3, Canada
*bilal.farooq@torontomu.ca

## ABSTRACT

There is an increased interest from transit agencies to replace fixed-route transit services with on-demand public transits (ODT). However, it is still unclear when and where such a service is efficient and sustainable. To this end, we provide a comprehensive framework for assessing the sustainability of ODT systems from the perspective of overall efficiency, environmental footprint, and social equity and inclusion. The proposed framework is illustrated by applying it to the Town of Innisfil, Ontario, where an ODT system has been implemented since 2017. It can be concluded that when there is adequate supply and no surge pricing, crowdsourced ODTs are the most cost-effective transit system when the demand is below 3.37 riders/$km^2$/day. With surge pricing applied to crowdsourced ODTs, hybrid systems become the most cost-effective transit solution when demand ranges between 1.18 and 3.37 riders/$km^2$/day. The use of private vehicles is more environmentally sustainable than providing public transit service at all demand levels below 3.37 riders/$km^2$/day. However, the electrification of the public transit fleet along with optimized charging strategies can reduce total yearly GHG emissions by more than 98%. Furthermore, transit systems have similar equity distributions for waiting and in-vehicle travel times.

## Main

It has become apparent that providing fixed-route public transit (FRT) is not always efficient and cost-effective, particularly in low-density areas[1–5]. Transit operators in such settings tend to reduce frequency and operating hours as well as adjust the coverage area to increase the occupancy of vehicles and reduce their operating costs. These circumstances can adversely affect the performance of the public transit system, leading to a greater reliance on private vehicles[5–7]. However, with the rapid advancements in information and communication technology, several on-demand public transit (ODT) systems have emerged as innovative solutions for low-density areas, such as demand responsive transit, ridesourcing, and mobility on-demand transit services[8–11]. ODT systems can be classified into three types based on the supply resources they use: dedicated fleet, crowdsourced, and hybrid ODT[8]. In dedicated fleet ODTs, the fare is primarily fixed, the fleet size remains stable, operating vehicles are often owned by organizations (e.g., municipalities), and drivers receive a fixed salary. Crowdsourced ODTs are subsidized ridesourcing services in which vehicles are privately owned, sharing is limited, and the fleet size varies based on the participation interest of private drivers. The third system, hybrid ODT systems, combines a dedicated fleet and crowdsourced ODT services.

In recent years, many public transit agencies around the world have launched or replaced their FRT service with an ODT. There are, however, several instances where the adoption of ODT services was not suitable or cost-effective, and the services had to be suspended[12–15]—examples include the dedicated fleet ODT in Helsinki, Finland, and Boston, Massachusetts, as well as the crowdsourced ODT in Manhattan and Brooklyn, New York[12]. Therefore, understanding where and when each ODT type is most efficient and cost-effective is essential for their sustainable adoption and operation.

The key elements of a sustainable transport system include being accessible, affordable, safe, secure, environmentally responsible, cost-efficient, and socially equitable[16–19]. Using simulation, prior studies have shown that ODT services can provide better service quality than the FRT in low-to-medium transport demand areas[10,20–23]. The replacement can significantly reduce walking as well as total travel time for passengers[10,12]. This is critical for passengers travelling in harsh weather conditions, at night, and in unsafe neighbourhoods. Other studies reported the reduction of total distance travelled[21,23], operating costs[23], and greenhouse gas (GHG) emissions[24]. There is, however, a wide variation in the literature with regard to the critical demand level at which the FRT becomes more efficient[10,20]. For example, a recent study compared the performance of the two services on grid and star networks. The results showed that the dedicated fleet ODT outperforms the FRT in terms of total travel time for demand up to 60 passengers/$km^2$/hour[10]. In another study, the two services were compared on a rectangular service area, using only one operating vehicle, and based on the weighed sum of the walking, waiting, and in-vehicle times of passengers. The results revealed that the critical demand density lies between 3.86 and 19.3 passengers/$km^2$/hour, depending on how the indicators are weighted[20].

Studies have suggested that crowdsourced ODT can also be effective in low-density areas, especially those with no or limited transport services[9,25,26]. A recent study examined the possibility of using a crowdsourced ODT to supplement the regional transit service in suburbs, revealing that the crowdsourced ODT reduces GHG emissions and travel costs when compared to individuals driving their own vehicles for the same trip[25]. A similar study found that the integration of crowdsourced ODT and public transport system can increase the demand for public transit as well as reduce traffic congestion and GHG emissions when compared to a scenario in which crowdsourced ODT is used exclusively[27]. Another recent study showed that the crowdsourced ODT service is more cost-effective than the dedicated fleet service when transport demand is limited[22]. In contrast, the service was found to increase operating costs[28] as well as cause traffic congestion[29] when modelled to replace existing transit systems in high demand areas. Yet, the critical demand density for each type of ODT remains unclear.

To ensure sustainable adoption and operation of ODT, it is also important to understand how they impact social equity and inclusion. Scholars indicate that ODT can theoretically contribute to social equity and inclusion in low-density areas[30–33]. Providing flexibility in routing and scheduling can facilitate travel for disadvantaged groups[34] and improve mobility for everyone[35]. A recent study evaluated accessibility for ridesourcing services in Austin, Texas, by modelling the waiting time using socioeconomic and built environment factors. The results revealed that waiting times were lower in low-income areas, but higher in areas with a high percentage of Hispanic/Latino and Black residents[36]. Another recent study assessed the usage distribution of exclusive and shared ridesourcing services among communities in Chicago. The results showed that under-served communities were positively correlated with shared rides and negatively correlated with exclusive rides[37]. However, it should be noted that there is an apparent lack of studies that have assessed the equity and inclusion aspects of ODT systems.

In this study, we present a comprehensive framework aimed at assessing the sustainability of ODT systems. The proposed framework contributes to the existing literature in three significant ways. Firstly, the framework evaluates the sustainability of ODT systems from multiple perspectives encompassing overall efficiency, environmental footprint, and social equity and inclusion. While some of these aspects have been explored individually in previous studies for various transportation systems, to the best of the authors' knowledge, no prior research has examined all these dimensions collectively in the context of ODT systems. Secondly, alongside traditional performance metrics, we propose the use of the generalized cost value, which is based on the generalized cost approach, as a new performance metric for ODT systems. This metric considers various aspects such as service cost, travel time, reliability, quality of service, and user preferences. By integrating these indicators, the generalized cost value offers a holistic metric to evaluate the overall efficiency of ODT systems, identify demand switching points between different ODT configurations, and quantify uncertainties within the performance evaluation of ODT systems. Thirdly, we provide a practical illustration of the proposed framework by applying it to a real transportation network (i.e., Town of Innisfil, Ontario), where a crowdsourced ODT system has been implemented since 2017. Through this case study, we provide valuable insights and recommendations for transit agencies and policy makers when planning new ODT systems. This empirical illustration offers strong evidence of the usefulness of the proposed metric in terms of invaluable insights and recommendations concerning the development and implementation of sustainable ODT systems.

## Methods

### Sustainability Analysis Framework for ODT Systems

Fig. 1 illustrates a general framework for assessing the sustainability of ODT systems from the perspective of overall efficiency, environmental footprint, and social equity and inclusion. Simulation models can be used to evaluate and compare the performance of different ODT systems within a study area without incurring full implementation costs and risks. Several inputs are required to model ODT systems, including transportation network data, network traffic demand, as well as public transit characteristics and demand data. The inputs can either be derived from a hypothetical or real-world transportation network. However, modelling a real-world transportation network using its actual characteristics and demand data can provide more reliable and accurate results, as the performance of ODT systems is highly influenced by the number of ride requests a system receives and their distribution over time and space[10]. Various simulation scenarios can be developed for each ODT system by changing its operational characteristics. Coverage area, supply settings (i.e., fleet size and vehicle capacity), and the interaction between supply and demand (i.e., the dispatching algorithm and its parameters) are the main factors that influence the performance of ODT systems. Additionally, crowdsourced ODT systems provide both shared and exclusive rides; therefore, the type of service can also be taken into account when modelling the system. The performance of simulated ODT configurations can be evaluated and compared using several performance indicators, which can be divided into indicators related to user performance and indicators related to system efficiency. User performance indicators include average waiting time, walking time, in-vehicle travel time, and trip length, while indicators of system efficiency include unmet demand, average occupancy, operator costs, average vehicle kilometres travelled per passenger, and GHG emissions.

Performance indicators are used to examine the sustainability of ODT systems in terms of their overall efficiency, environmental footprint, as well as social equity and distribution. In our study, we propose the use of the generalized cost value, which is based on the generalized cost approach, as a new performance metric for ODT systems. This metric combines several important indicators, including user total travel time, operator costs, served demand, and value of time. In order to convert the total travel time of riders into a monetary equivalent, the value of time is used[22,38]. By integrating these indicators, the generalized cost value offers a comprehensive and holistic evaluation of the overall efficiency of ODT systems, considering various aspects such as service cost, travel time, reliability (waiting time), quality of service (served demand), and user preferences (value of time). Moreover, the new metric provides decision-makers with a more thorough understanding of the trade-offs and impacts associated with different configurations of a system, which in turn helps them to make informed decisions. By comparing generalized cost values for ODT systems over a range of demand levels, critical demand levels can be identified.

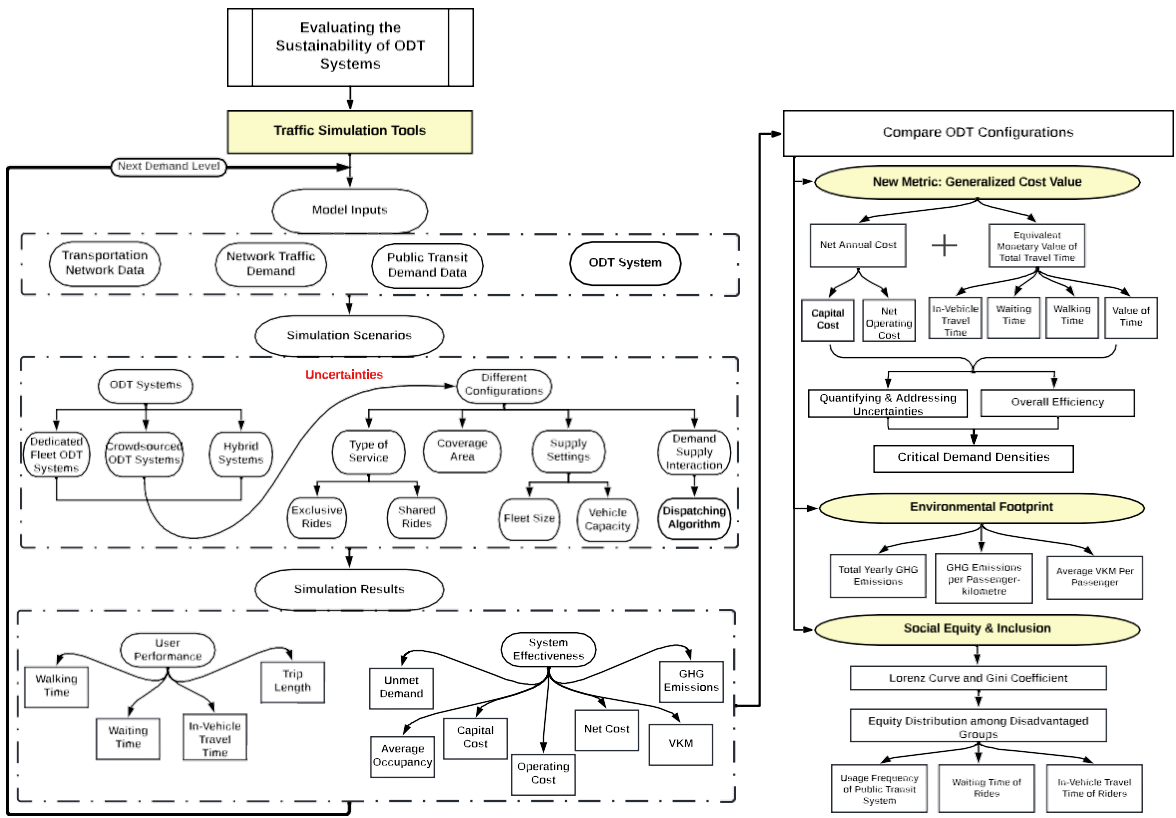

**Figure 1.** General framework for assessing the sustainability of ODT systems.

The environmental footprint of ODT systems can be assessed using three indicators; the average vehicle kilometres traveled per passenger, total yearly GHG emissions, and GHG emissions per passenger-kilometre. Finally, Lorenz curve and Gini coefficient are proposed to evaluate the social inclusion and equity of ODT configurations. Both Lorenz curve and Gini index were initially used in economics to measure the income inequality across a nation or a social group. However, they have been widely used to assess equity and inclusion aspects of transportation services. The Lorenz curve illustrates the (in)equality of a distribution graphically, while the Gini coefficient measures the level of (in)equality mathematically. Gini coefficient values range from 0 to 1, where 0 indicates perfect equity and 1 indicates perfect inequity. Generally, the larger the value, the greater the inequity[39,40]. In this framework, Lorenz curve and Gini coefficient methods are proposed to examine the equity distribution in public transit usage within a study area as well as equity distribution in waiting times and in-vehicle travel times for simulated ODT configurations among disadvantaged populations.

It is important to note that transportation systems are subjected to a wide range of uncertainties, such as fluctuations in demand, variations in supply, variations in travel times, and disruptions in operations. Taking into account these uncertainties in the modelling framework allows us to simulate and analyze the performance of ODT systems under realistic conditions. It helps us understand how ODT systems respond to different scenarios; therefore, developing proactive strategies to deal with uncertainty. The proposed framework considers uncertainty and heterogeneity by evaluating ODT configurations across a wide range of demand levels, setting up varying supply conditions for each ODT system, modelling different ODT system types, and considering different service types for crowdsourced ODT systems. The considerations outlined above allow for a more realistic representation of real-world complexities and variations, which enhances the reliability and applicability of the analysis. On the other hand, the generalized cost value provides a robust framework for capturing and addressing uncertainties within the performance evaluation of ODT systems by including various factors into the cost calculation, including variability in travel time, demand fluctuations, disruptions in service, and user preferences. Therefore, the inclusion of the generalized cost value not only allows a comprehensive evaluation of the performance of ODT systems, but also provides a valuable tool for quantifying and addressing uncertainties within the modelling framework.

## Case Study: Town of Innisfil, Ontario

The proposed framework has been applied to Innisfil, where a crowdsourced ODT system has been implemented since 2017. The availability of actual ODT operational data for Innisfil presents a unique opportunity to assess the sustainability of different ODT systems using actual transportation network data, real demand patterns, and observed demand levels. Towards this end, we have developed a micro-simulation model using the open-source platform SUMO, Simulation of Urban Mobility[41].

SUMO is an open-source, microscopic traffic simulation model that has been selected for this study due to its availability and wide adoption in modelling different aspects of urban transportation. Its ability to simulate individual vehicle behaviour allows us to measure the impacts of traffic demand and assignment on the performance metrics of ODT systems. Additionally, SUMO can facilitate the quantification of uncertainties in traffic dynamics, although for consistency, the same dataset was used across all scenarios in this study.

We simulate the transportation network of Innisfil under various ODT design configurations. The model is sought to evaluate the overall efficiency of several ODT configurations, identify demand switching points between them, and evaluate the environmental footprint, social equity, and inclusion of each configuration. To ensure the reliability of our findings, the simulation model has been calibrated based on the existing ODT system in Innisfil, using actual operational data and real transportation network data.

By utilizing operational data, actual transportation network data, and a calibrated simulation model, this study provides a reliable assessment of the sustainability of different ODT configurations. Therefore, the findings are expected to provide transit agencies and policymakers with evidence-based guidance and recommendations for the design and implementation of sustainable ODT systems in similar settings.

### *Overview*

Innisfil is a small municipality in Ontario located on the northern edge of the Greater Toronto and Hamilton Area (GTHA). It has a population of 36,566 residents, a land area of 262.4 square kilometres, and therefore a population density of 139 residents/$km^2$[42]. In May 2017, the Town partnered with Uber to provide a subsidized crowdsourced ODT service to the residents instead of investing in a new FRT system. The subsidization of the transit service entails discounts on trips to and from the Town, as well as flat-fares for many predefined places within and around it[26,43].

### *Model inputs and calibration*

The transportation network of Innisfil was extracted from OpenStreetMap[44]. Innisfil provided the actual demand data for the crowdsourced ODT (Uber Transit) during the COVID-19 pandemic. The dataset contained detailed information on trip request time, pickup and drop-off times and locations, as well as trip length, duration, and waiting time. The average daily traffic value for Innisfil was utilized from the 2016 Transportation Tomorrow Survey (TTS)[45].

We first simulated the existing crowdsourced ODT in Innisfil using March 15, 2021, ride requests, which represent the average weekday demand during the COVID-19 pandemic. The aim of this step was to calibrate the model by estimating the spatiotemporal distribution of the supply that corresponds to the actual situation. Ride requests were processed as exclusive rides using the Greedy heuristic dispatching algorithm, as the option of ridesharing was not permitted in March 2021. The Greedy algorithm is a predefined dispatching algorithm in SUMO that assigns vehicles to the nearest waiting passengers based on the First-Come-First-Served (FCFS) approach. While the simulation day represents demand during the pandemic, the origin-destination patterns of the demand are similar to the pre-pandemic patterns but with reduced demand levels[43]. Therefore, the pandemic effects on the demand were taken into account in the later scenarios by increasing the demand level and enabling the ridesharing option.

Traffic flows were generated and distributed randomly based on the network's average daily traffic value (55,000 trips/day). Since Innisfil has an uncongested network, the flow of trips was simulated using the distance-based shortest path algorithm. The simulation results were verified using the paired t-test method. The comparison showed no statistical difference between the simulation results and the actual data, in terms of in-vehicle travel times, trip length, and waiting time values, at 95% confidence level (Supplementary Table 1).

### *Simulation scenarios*

In total, nine ODT scenarios were developed and simulated separately. The simulations covered a 24-hour period, and the demand in all scenarios ranged from 50% to 500% of March 15, 2021 (177 riders/day), the base-demand level, with increments of 50%. Moreover, the same passenger and traffic demand datasets were used across the simulated scenarios to ensure consistency. The first four scenarios simulated the existing crowdsourced ODT service with and without the ridesharing option (Fig. 2a). In both cases, the spatiotemporal distribution of supply resulting from the calibration was used for the base demand level (100%). After that, we set up two scenarios for each case to account for possible variations in supply in real-life. In one scenario, the supply remained constant while demand decreased or increased ($α = 0$), representing the worst-case scenario, where there is a shortage of supply. In the other scenario, the supply corresponded the demand linearly with a slope of 1 ($α = 1$), representing the best-case scenario where there is always enough supply to satisfy demand.

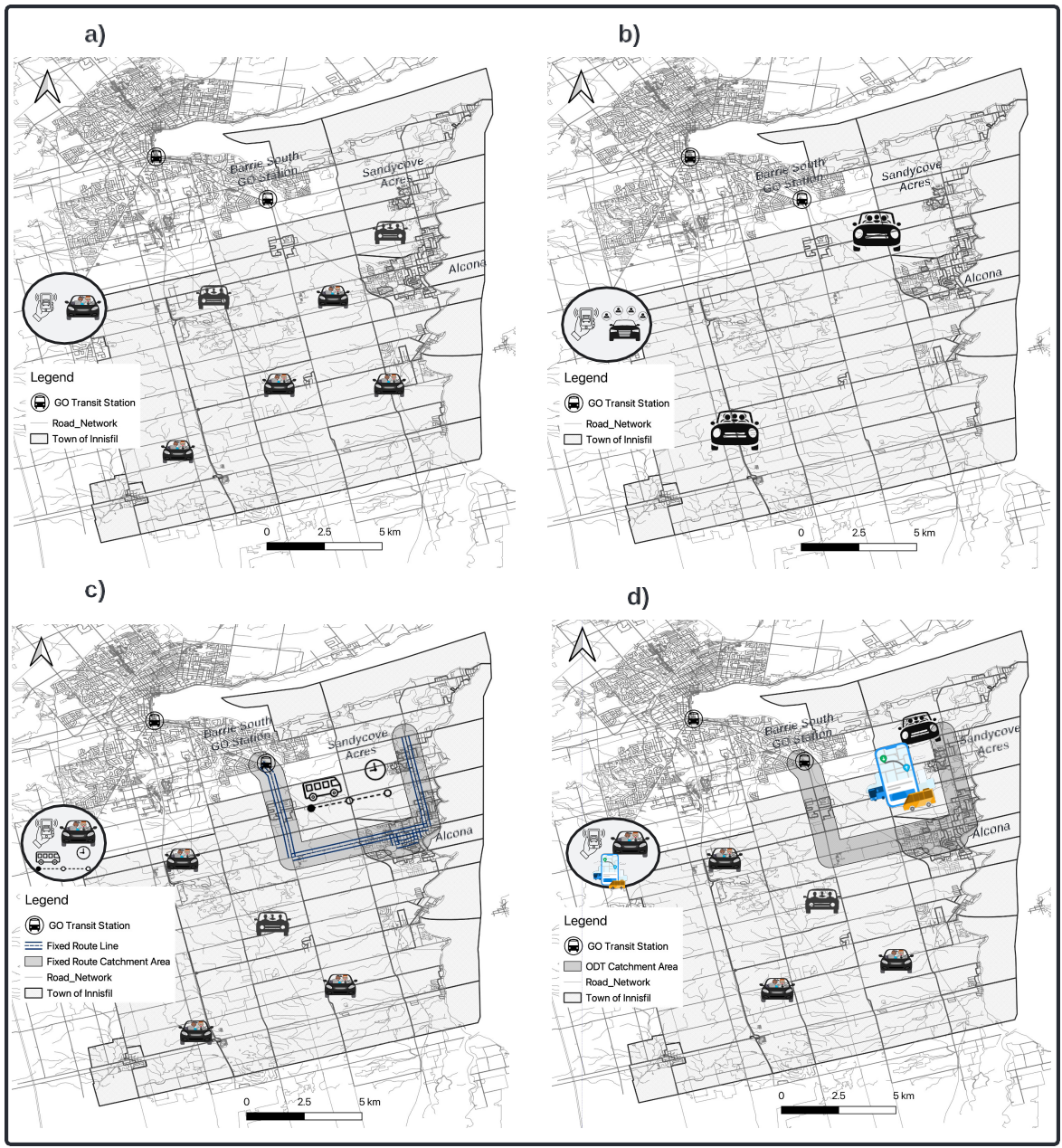

**Figure 2. Main public transit types used in this study.** a) Crowdsourced ODT service. b) Dedicated fleet ODT service. c) Hybrid transit system: Dedicated fleet FRT and crowdsourced ODT services (FRT-based hybrid system). d) Hybrid transit system: Dedicated fleet and crowdsourced ODT services (ODT-based hybrid system). This figure was generated using the free and open-source software QGIS under the CC BY 4.0 license.

The Greedy algorithm was used for exclusive ride scenarios, and the Shared Greedy algorithm for ridesharing scenarios.

Shared Greedy is a modified version of Greedy, in which another waiting passenger is picked up while on the way to the first passenger's destination, provided the maximum detour limit is not exceeded. This study adopted the Shared Greedy algorithm with a maximum detour factor of two. Although more advanced algorithms, such as dynamic programming, may enhance the efficiency of the dispatching process by grouping passengers with similar itineraries, their benefits are greater in congested and high-demand networks. The transportation network of Innisfil is characterized by an uncongested nature, low public transit demand, and a large land area. Given the low probability of having many ride requests with similar itineraries in such settings, the performance of Greedy algorithms, which is computationally more efficient and easier to implement, is expected to be comparable to more complex algorithms in this context.

The next three scenarios involved modelling a dedicated fleet ODT service (Fig. 2b). We modelled this service using SUMO's demand-responsive transit (DRT) tool, which allows multiple passengers to be served simultaneously by a single vehicle. The tool aims to maximize the number of requests served and minimize fleet mileage while satisfying some constraints, such as maximum detour factor and maximum waiting time value. SUMO's DRT tool is a robust simulation tool designed primarily for modelling and analyzing shared DRT systems. It uses Traffic Control Interface (TraCI) and the functionality of the taxi device to control requests and operate the fleet effectively. The DRT tool uses a dispatcher to dynamically manage incoming requests. The dispatcher aims to combine multiple requests to maximize the number of served demand while minimizing the total distance travelled by the fleet[46]. There are two primary components of the DRT tool: the scheduling module and the Dial-A-Ride Problem (DARP) solver. TraCI is used to orchestrate the simulation and oversee the fleet of vehicles and DRT demand. The scheduling module detects new and pending requests and triggers the DARP solver for finding the most optimal routes for each vehicle[46]. For further details about SUMO's DRT tool we refer readers to [46] and the official SUMO's website at: https://shorturl.at/hiuyE.

The service was first simulated using the March 15th, 2021, ride requests in order to determine the optimal supply for the base demand level. The optimal supply was defined as the minimum fleet size needed to meet all trip requests. We then created three scenarios for the dedicated fleet ODT service. In one, the supply remained constant while demand decreased or increased ($α = 0$), while in the other two, the supply followed the demand linearly with slopes of 0.5 ($α = 0.5$) and 1 ($α = 1$). This is mainly for identifying the most cost-efficient supply setting as demand rises, since the dedicated fleet ODT system requires the purchase of public vehicles. We used a maximum detour factor of two, and a maximum waiting time of thirty minutes for all dedicated fleet ODT scenarios.

Lastly, two hybrid transit scenarios were created. One scenario involved a hybrid transit system combining dedicated fleet FRT with crowdsourced ODT services (FRT-based hybrid system). The service area of the FRT was identified based on an analysis of the actual origin-destination (OD) flows of the existing crowdsourced ODT service. The analysis yielded four potential routes, but the one with the highest performance and demand was chosen (Fig. 2c). It operated from 7:00 AM to 9:00 PM, covering a length of 20 kilometers and connecting three key locations: (a) a large residential area in the northeastern part of the Town (Sandycove Acres), (b) the central area of the Town (Alcona), and (c) the regional transit station in the City of Barrie (Barrie South GO station). Moreover, the FRT was designed to serve passengers within a seven-minute walking distance, with both pickup and drop-off locations within its service area. The crowdsourced ODT, on the other hand, covered the rest of the Town during the FRT operating hours, and the entire Town outside of those hours. In the same way as the previous scenarios, we estimated the optimal fleet size for each service based on the base level of demand. Two vehicles were used for the dedicated fleet FRT service for demand less than 300% of the simulated day, and three vehicles thereafter. As for the crowdsourced ODT service, the supply was assumed to follow the demand linearly with a slope of 1 ($α = 1$). The other scenario involved a hybrid transit system combining dedicated fleet ODT with crowdsourced ODT services (ODT-based hybrid system). The dedicated fleet ODT was designed to cover the same service area and hours as the FRT in the previous scenario with a maximum detour factor of two, and a maximum waiting time of thirty minutes (Fig. 2d). In both services, the supply was set to follow the demand linearly with a slope of 0.5 ($α = 0.5$) for the dedicated fleet service, and 1 for the crowdsourced service ($α = 1$).

$$S = α \times D \tag{1}$$

where $S$ is the percent change in supply compared to the base level, $D$ is the percent change in demand compared to the base level, and $α$ is the slope. We set the $α$ values to 0 and 1 for the crowdsourced ODT scenarios, and to 0, 0.5, and 1 for the dedicated fleet ODT scenarios. For the hybrid scenarios, $α$ was set to 1 for the crowdsourced ODT service and 0.5 for the dedicated fleet ODT service.

To ensure consistency across all scenarios, the decision was made to use eight-seat vehicles, as determined by the findings of[25]. Although the restriction to a specific vehicle type may limit the exploration of supply-side variations, it allowed for consistent comparison of ODT configurations. Future research could focus on investigating and optimizing supply-side features, including vehicle type. Supplementary Fig. 1 presents the temporal distribution of demand and supply for crowdsourced, dedicated fleet, and hybrid ODT systems at the base level of demand. Furthermore, the differences between the dispatching algorithms used in the crowdsourced ODT and the dedicated fleet ODT scenarios are illustrated in Supplementary Fig. 2.

*Performance indicators*

Scenarios were evaluated using a set of performance indicators, including walking time, waiting time, in-vehicle travel time, trip length, served demand, and net annual costs. The net annual costs for each scenario ($NAC$) were calculated as the sum of capital ($CC$) and net operating costs ($NOC$).

$$NAC = CC + NOC \tag{2}$$

For the crowdsourced ODT scenarios, the capital cost component was set to zero, as the operating vehicles are privately owned. The net operating costs ($NOC_{CS}$), on the other hand, were calculated by multiplying the average cost per trip by the annual demand served ($SD$). The average cost per trip included fixed fees (base and booking fees), a ride time charge ($\beta_{time}$) multiplied by the average in-vehicle time ($IVTT$), and a distance charge ($\beta_{length}$) multiplied by the average trip length ($TL$), minus the fare. According to the fare structure listed on Uber's website, the fixed fees for UberX service (exclusive rides) in the GTHA is 5.25 CAD, the ride time charge is 0.18 CAD per minute, and the distance charge is 0.81 CAD per kilometer. The same rates apply to UberPool service (shared rides) except that the fixed fees are lower at 4.25 CAD[47]. Moreover, a flat fare of 4 CAD was assumed for all scenarios.

$$NOC_{CS} = (FixedFees + \beta_{time} \times IVTT + \beta_{length} \times TL - Fare) \times SD \times 365 \tag{3}$$

For the dedicated fleet ODT and FRT services, the capital costs were calculated by the number ($N$) and unit price ($V_{Price}$) of the vehicles involved. Toyota Sienna 2022 vehicles were assumed to be used for the dedicated fleet services. According to the official website of Toyota Canada, the model is priced at 41,050 CAD.

$$CC = N \times V_{Price} \tag{4}$$

The net operating costs for the dedicated fleet ODT ($NOC_{DF}$) scenarios were calculated as follows:

$$NOC_{DF} = (OC \times Avg_{Vehicles} \times OH - F \times SD) \times 365 + OtherCosts \tag{5}$$

where $OC$ is the average operating cost of the ODT vehicle per hour, including operators' wages, vehicle fuel, maintenance, and insurance costs, admin, booking dispatch, and reporting. $Avg_{Vehicles}$ is the average number of operating vehicles, $OH$ is the daily hours of operation, $F$ is the fare, $SD$ is the served demand, and $OtherCosts$ is the additional costs associated with the service, including wages for supervisors, office supplies, and so on. The average operating cost per vehicle per hour was set at 83.95 CAD, based on publicly available data for similar services in Ontario[48] and Alberta[49]. Moreover, we set 200,000 CAD for the additional expenses of the service, assuming one supervisor and one customer service representative can manage it.

As for the dedicated fleet FRT ($NOC_{FRT}$) scenarios, the net operating costs for were calculated as follows:

$$NOC_{FRT} = (OC_{km} \times N \times VKM + N \times OH \times W - F \times SD) \times 365 + OtherCosts \tag{6}$$

where $OC_{km}$ is the average operating cost of the FRT vehicle per kilometer, including vehicle fuel, maintenance, and insurance costs. $N$ is the number of operating vehicles, $VKM$ is the vehicle kilometres travelled, $OH$ is the daily hours of operation, $W$ is the hourly wage for drivers, $F$ is the fare, $SD$ is the served demand, and $OtherCosts$ is the additional costs associated with the service. The average operating cost per vehicle per kilometer was set at 0.73 CAD based on an online tool that uses existing vehicle data in Canada[50]. The cost of drivers was calculated using the minimum wage in Ontario, 15 CAD per hour[51]. The additional expenses of the service were assumed to be 200,000 CAD as in the dedicated fleet ODT scenarios.

*Environmental footprint*

We used vehicle kilometres travelled per passenger and GHG emissions indicators to assess the environmental footprint of the simulated transit systems. Transit systems were compared both to each other and to a baseline scenario in which demand would be met by private single occupancy vehicles. In order to ensure comparability, we assumed that the fleet in the crowdsourced ODT scenarios would consist of the same vehicle type as the dedicated fleet scenarios. On the other hand, we assumed that the private vehicles in the baseline scenario would be Honda Civics, which are among the best-selling vehicles in Canada[52]. In both cases, the 2019 model was considered, as not all vehicles in Innisfil are brand new or old. The annual GHG emissions ($Total_{GHG}$) for all scenarios were calculated as follows:

$$Total_{GHG} = GHG_{km} \times Total_{km} \times 365 \tag{7}$$

where $GHG_{km}$ is the average annual GHG emissions per vehicle-kilometre in tons, and $Total_{km}$ is the total daily kilometres travelled by the fleet. The average annual GHG emissions per vehicle-kilometre for each vehicle type was calculated using an online tool[53].

GHG emissions were calculated under the most conservative scenario, where all vehicles were gasoline-powered. However, the environmental footprint of the simulated transit systems was also examined under five levels of electrification, from 20% to 100% with increments of 20%. Electric vehicles emit no GHGs on the road, but there are indirect emissions associated with the production of electricity needed to charge them[54]. The indirect GHG emissions from electric vehicles ($EV_{GHG}$) at each electrification level were calculated as follows:

$$EV_{GHG} = I \times E \times Total_{km} \times 365 \tag{8}$$

where $I$ is the GHG intensity of Ontario's electricity grid, was 25 grams of GHGs per kilowatt-hour electricity generated in 2020[55]. $E$ is the average energy consumption of an electric vehicle per kilometre, for which, we used the values from Dynamometer Database [56]. $Total_{km}$ is the total daily kilometres travelled by the fleet. To simplify the analysis, we assumed that the adoption of electric vehicles would not alter the fleet size or total kilometers travelled of the simulated transit systems.

### *Critical demand densities*
The generalized cost approach to assess the overall efficiency of the simulated transit systems and identify the demand switching points among them. The value of time for passengers, in this study, was assumed to be 15 CAD per hour, which is the Ontario minimum wage.

$$GC = \frac{(WK_T + WT_T + IVTT)}{60} \times SD \times 365 \times VOT + NAC \tag{9}$$

where $GC$ is the generalized cost of a transit system in CAD, $WK_T$ is the average walking time in minutes, $WT_T$ is the average waiting time in minutes, $IVTT$ is the average in-vehicle travel time in minutes, $SD$ is the daily served demand, $VOT$ is the value of time for passengers expressed in CAD per hour, and $NAC$ is the service net annual cost in CAD.

### *Social inclusion & equity*
Using the Lorenz curve and the Gini coefficient, we first evaluated the equity distribution in the current use of Innisfil's crowdsourced ODT service among disadvantaged communities with data from March 15, 2021. We then examined the equity distribution in waiting times and in-vehicle travel times for simulated ODT configurations among disadvantaged populations at the base and the maximum demand levels, 177 riders/day and 885 riders/day, respectively. We performed both analyses at the level of the dissemination area, which is the smallest geographical area in Census zoning system[57]. Disadvantaged communities considered in this study included neighbourhoods with low population density and with high percentages of seniors, low-income, young adults, unemployed, single parents, and low-educated residents. Although other groups, such as carless, people with a disability, newly arrived migrants, and minority ethnic people, could encounter transport-related social exclusion issues, only limited information about these groups can be accessed from Census data.

## Results

Here, we first present the simulation results in terms of user performance, operator's costs, and environmental footprint. Next, we evaluate the overall efficiency of the simulated transit systems and identify the demand switching points. Finally, we evaluate the social equity and inclusion of the simulated transit systems.

### User performance and operator's costs
Fig. 3a presents the key performance measures for passengers. The performance of transit systems is comparable when the demand is at or below the base-demand level; however, as the demand goes up, the differences become more apparent. The uncongested nature of Innisfil's network ensures consistent performance of transit systems, in terms of waiting time, in-vehicle time, and trip length, in scenarios where sufficient supply is available across all levels of demand. When the supply remains constant, however, the performance deteriorates with increasing demand.

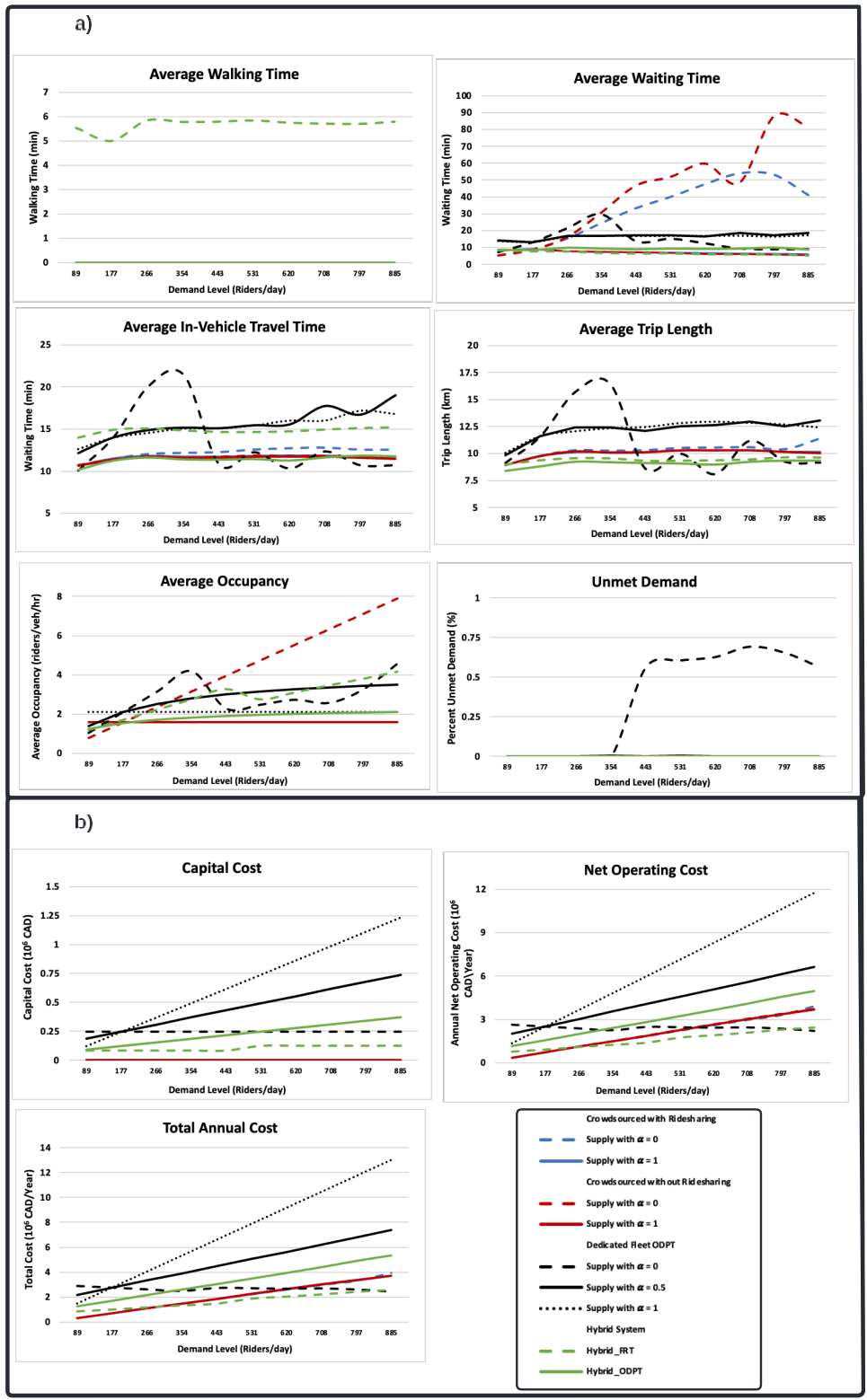

**Figure 3. Key performance measures of transit systems.** a) User performance measures. b) Operator's costs.

On average, passengers walk six minutes to reach the nearest stop in the FRT-based hybrid system. For all other transit systems (ODT-based hybrid system, dedicated fleet ODTs, and crowdsourced ODTs), walking time is assumed to be zero, as they provide door-to-door service. Across all levels of demand, the lowest waiting and in-vehicle times, as well as the shortest trip lengths are observed with ODT-based hybrid system and crowdsourced ODTs with increased supply. On the other hand, the highest occupancy rates and the longest waiting times are observed with dedicated fleet ODTs when the demand is at or below 354 riders/day, and crowdsourced ODTs with constant supply when the demand exceeds 354 riders/day. Moreover, dedicated fleet ODTs have the longest in-vehicle times and trips lengths at all demand levels. The results can be explained by considering the differences in operational characteristics between the dedicated fleet and crowdsourced ODTs.

In dedicated fleet ODTs, multiple requests can be served simultaneously by a single vehicle provided that both waiting time and detour factor constraints are met. Ride requests are rejected if neither constraint can be met by the operating vehicles. On the other hand, in crowdsourced ODTs, there are no waiting time constraints, vehicles can serve a maximum of two requests at once, and ride requests remain in the system until they are met. In light of this, dedicated fleet ODTs provide a higher level of sharing, which increases detours and, therefore, passengers in-vehicle time and trip length. Moreover, the dedicated fleet ODT with constant supply reaches its capacity at 354 riders/day. The system fails to accommodate most ride requests after this point, resulting in a sharp decrease in waiting times, in-vehicle times, trip lengths, and occupancy rates. In contrast, crowdsourced ODTs accommodate all ride requests regardless of demand. However, when supply is constant, high demand results in extremely long waiting times. With this in mind, it is important to implement an incentive strategy for the drivers to maintain an adequate supply of vehicles as demand increases.

The capital, operating, and total costs of the simulated transit systems are compared in Fig. 3b. It can be seen that crowdsourced ODTs have no capital costs since the operating vehicles are privately owned. In contrast, the highest capital costs are associated with dedicated fleet ODTs, since they require the purchase of public vehicles. The lowest annual operating costs are observed with crowdsourced ODTs when the demand is below 266 riders/day, and with the FRT-based hybrid system when the demand exceeds 266 riders/day. Moreover, crowdsourced ODTs have the lowest total annual net cost when the demand is below 292 riders/day, and the FRT-based hybrid system when the demand is above 292 riders/day. Conversely, dedicated fleet ODTs have the highest operating and total costs at all demand levels.

Here, the user performance and operator costs are assessed independently. However, the generalized cost approach is used in next sections to assess the overall efficiency of simulated transit systems by combining both user travel time and operator costs.

**Environmental footprint**

We use vehicle kilometres travelled (VKM) per passenger and GHG emissions measures to assess the environmental footprint of the simulated transit systems. In Fig. 4, transit systems are compared both to each other and to a baseline scenario in which demand is met by private single occupancy vehicles. In all scenarios except the one where crowdsourced ODT operates without ridesharing and with constant supply, as demand increases, VKM per passenger and GHG emissions per passenger-kilometre decrease, while yearly GHG emissions increase. This is because as demand increases, there is an increased chance of finding requests closer to the onboard passenger's destination, thus reducing deadheading distance. Furthermore, in ridesharing scenarios, more requests can be combined and assigned to a single vehicle as demand increases. Serving more requests, however, leads to more kilometres travelled by the fleet and, thus higher GHG emissions. On the other hand, in the case of exclusive rides and constant supply, the average VKM per passenger increases as demand increases. Crowdsourced ODTs process ride requests on a First-Come, First Served (FCFS) basis. Thus, when supply is limited, idle vehicles may have to travel long distances to pick up remote requests instead of nearby ones.

In comparison with other transit systems, the dedicated fleet ODT with constant supply has the lowest VKM per passenger and yearly GHG emissions at all demand levels, as it provides the highest level of sharing. Conversely, crowdsourced ODTs with constant supply are associated with the highest VKM per passenger and yearly GHG emissions. Contrary to expectations, simulated transit systems produce more GHG emissions than passengers driving their own vehicles (baseline scenario) at all demand levels. There are two possible reasons for this. First, transit systems cover a large area of 262 square kilometres, while the demand is not evenly distributed. Thus, vehicles make long deadheading trips and, in ridesharing scenarios, long detours. On the other hand, the baseline scenario does not involve deadheading trips or detours, as passengers drive directly from their origin to their destination. Second, we use larger vehicles for transit systems (eight-seaters), but these may not be large enough to provide the level of sharing the results in lower per passenger kilometre GHG emissions. However, it is observed that as demand increases, GHG emissions per passenger kilometre for dedicated fleet ODTs decrease and becomes closer the baseline scenario level. Thus, it is expected that as the Town grows and the network becomes congested, higher capacity public transit vehicles will likely result in lower GHG emissions per passenger than private vehicles.

**Critical demand densities**

The generalized costs of the simulated transit systems are compared in Fig. 5. For the dedicated ODT with constant supply, the generalized costs are calculated only for demand densities less 1.35 passengers/$km^2$/day, since the system cannot accommodate most ride requests above this level.

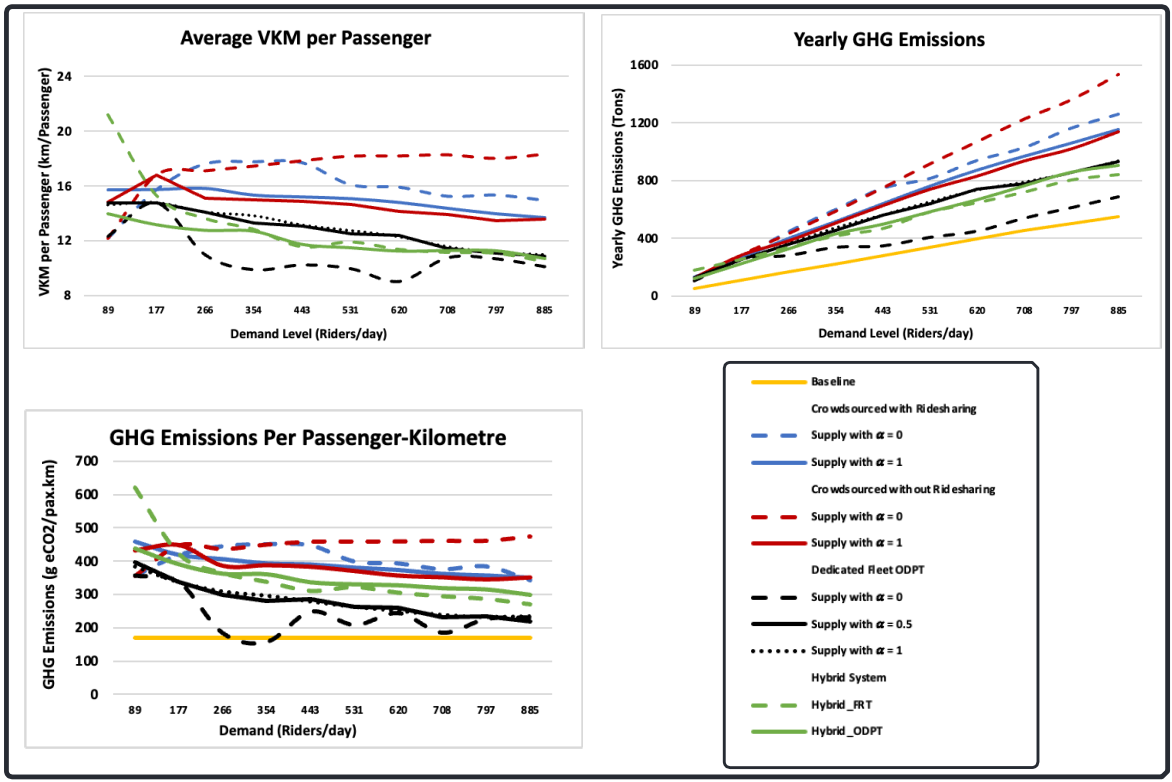

**Figure 4.** Transit systems environmental footprint measures.

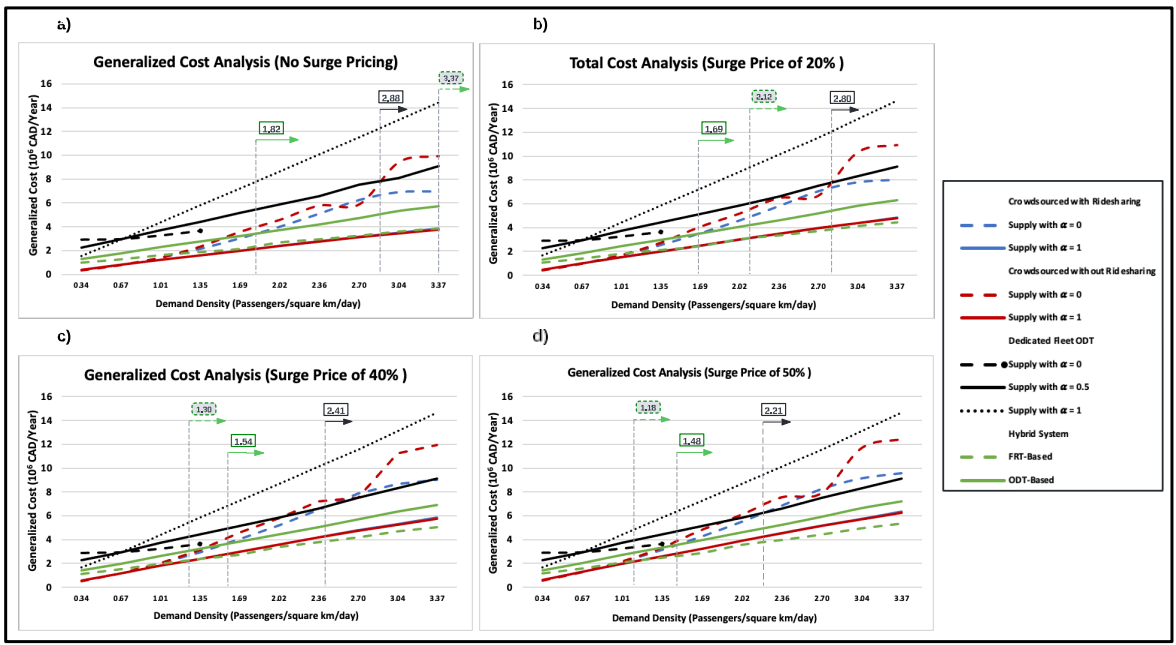

**Figure 5.** Generalized costs of the simulated transit systems.

Fig. 5a indicates that crowdsourced ODTs with increased supply are the most cost-efficient transit system for all demand levels below 3.37 riders/$km^2$/day (885 riders/day in the Innisfil context). However, both crowdsourced ODTs with increased supply and FRT-based hybrid systems have approximately the same generalized cost value when the demand is at 3.37

**11/20**

riders/$km^2$/day. The ODT-based hybrid system is more cost-efficient than crowdsourced ODTs with constant supply when demand exceeds 1.82 riders/$km^2$/day. Moreover, it is observed that when the demand exceeds 2.88 riders/$km^2$/day, the dedicated fleet ODT with increased supply ($α$=0.5) becomes more cost-efficient than the crowdsourced ODT with exclusive rides and constant supply.

The results shown in Fig. 5a do not take into account surge pricing, which occurs when demand for crowdsourced ODTs exceeds supply during specific hours and locations[58]. We introduced three cost increments on the operating costs of crowdsourced ODTs, at 20%, 40%, and 50%, in order to assess the impact of surge pricing on critical demand densities. The results are shown in Figs. 5 b, c, and d, respectively. Accordingly, it is observed that the critical demand density between crowdsourced ODTs and the FRT-based hybrid system decreases to 2.12, 1.30, and 1.18 riders/$km^2$/day, respectively. Furthermore, the critical demand density between the dedicated fleet ODT with increased supply ($α$=0.5) and crowdsourced ODTs with constant supply becomes 2.8, 2.41, and 2.21 riders per day, respectively.

## Social inclusion & equity

Lorenz curve and Gini coefficient are used at dissemination area level to (a) evaluated the equity distribution in the current use of Innisfil's crowdsourced ODT service among disadvantaged communities and (b) examine the equity distribution in waiting times and in-vehicle travel times for simulated ODT configurations among disadvantaged communities.

### *The current use of the crowdsourced ODT service*

According to Innisfil's annual travel satisfaction survey, 70% of users are satisfied with the existing crowdsourced ODT[59]. The service has four times higher accessibility levels than providing a FRT[60]. Moreover, assuming that the service has a flat fare of 4 CAD and based on Litman's definition of transportation affordability[61], 75% of the households in Innisfil can afford to make two trips per day.

Fig. 6 presents the equity distributions of the existing crowdsourced ODT in Innisfil among disadvantaged communities. It is observed that there are low rates of inequality in the use of the service across income, education, employment, young adulthood, seniors, and single parenthood. This indicates that the service is effective at meeting residents' travel needs on the one hand, as well as eliminating mobility barriers for disadvantaged communities on the other. However, lower trip production rates are observed in communities with the lowest incomes and lowest levels of education. Therefore, offering more subsidies and discounts to low-income residents can ensure their inclusion and further enhance the equity aspect of the service. Moreover, there is a high rate of inequality in the use of the service across population density levels. Higher trip production rates are associated with the low-density areas. Most likely, this is the result of riders returning from work, shopping, and large parks. This is in line with our recent findings indicating that neighbourhoods with industrial, commercial, and large parks in Innisfil are associated with higher demand for the crowdsourced ODT[43].

### *Waiting and in-vehicle travel times for ODT systems*

Fig. 7 presents the Gini coefficient values for the waiting and in-vehicle times for transit systems at the base (177 riders/day) and maximum demand levels (885 riders/day). Transit systems generally have similar equity distributions for waiting and in-vehicle travel times. In both demand levels, income is associated with the lowest inequality in waiting times and in-vehicle travel times, while population density is associated with the highest. At the base level of demand, there is a low inequality rate in waiting times and in-vehicle travel times across income. There are medium inequality rates across education, employment, young adulthood, seniors, and single parenthood, and a high inequality rate across population density. Compared to other transit systems, the FRT-based hybrid system exhibits the smallest rate of waiting time inequality across income, population density, single parenthood, and education, but the highest rate among seniors. The dedicated fleet ODT has the lowest inequality rate in waiting times among seniors, while the highest rates are found in population density, employment, single parenthood, and young adulthood. The crowdsourced ODT with exclusive rides has the lowest inequality rate across employment, but the most across income. The ODT-based hybrid system shows the smallest rate of waiting time inequality across young adulthood. In terms of in-vehicle travel time, the ODT-based hybrid system exhibits the lowest rates across income, population density, young adulthood, seniors, and employment. The dedicated fleet ODT, on the other hand, has the smallest rates across single parenthood and education. In contrast, the highest rates are observed for crowdsourced ODTs across income, education, seniors, and population density, the dedicated fleet ODT across employment and young adulthood, and the FRT-based hybrid system across single parenthood.

At the maximum demand level, there is a high inequality rate in waiting times and in-vehicle travel times across population density and employment. There are medium inequality rates across education, young adulthood, seniors, and single parenthood, and a low inequality rate across income. In comparison to other transit systems, the dedicated fleet ODT with increased supply ($α$ =0.5) has the lowest rate of inequality rate in waiting times among single parenthood, employment, and education. The FRT-based hybrid system has the lowest rate of waiting time inequality across income, but the highest rates across seniors and employment. The crowdsourced ODT with exclusive rides and increased supply has the lowest rate of waiting time inequality across seniors, but highest rates for population density, single parenthood, and education. Moreover, the crowdsourced ODT with exclusive rides and constant supply has the smallest rate of waiting time inequality across young adulthood, while the crowdsourced ODT with shared rides and constant supply has the smallest rate inequality across population density. On the flip side, the highest rates of waiting time inequality across income and young adulthood are observed for the ODT-based hybrid system. As for in-vehicle travel time, the FRT-based hybrid system shows the lowest rates across employment, single parenthood, and education. the lowest inequality rate in in-vehicle travel times across young adulthood is observed with the dedicated fleet ODT with increased supply ($α$ =1). The crowdsourced ODT with exclusive rides and constant supply has the lowest inequality rates across income and seniors, but the highest rates across population density, single parenthood, and education. On the other hand, the dedicated fleet ODT with increased supply ($α$ =0.5) has the lowest rate of inequality rate across population density, but highest rates across income, seniors, young adult hood, and employment.

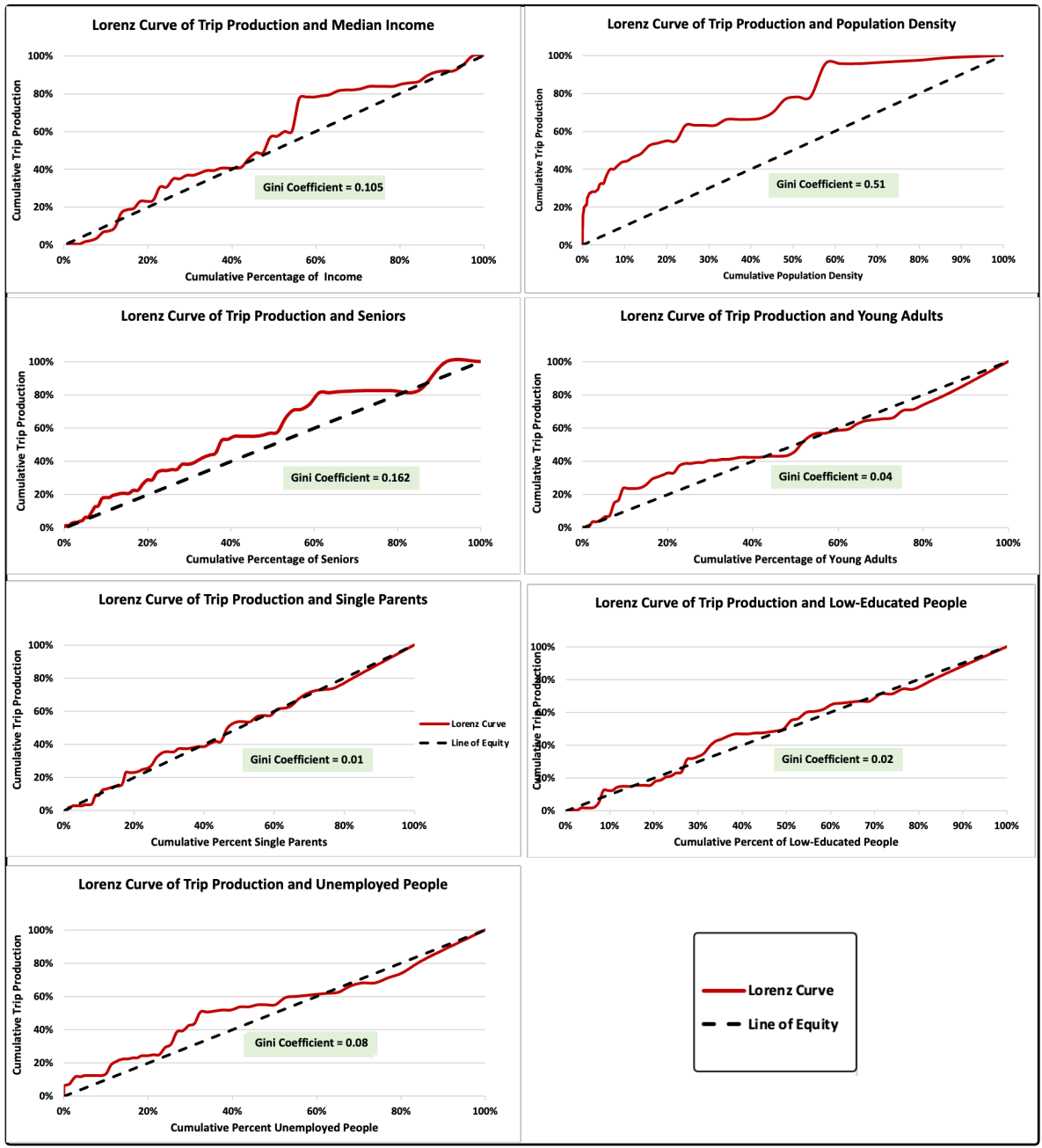

**Figure 6.** Equity distributions in the current use of Innisfil's crowdsourced ODT service among disadvantaged communities.

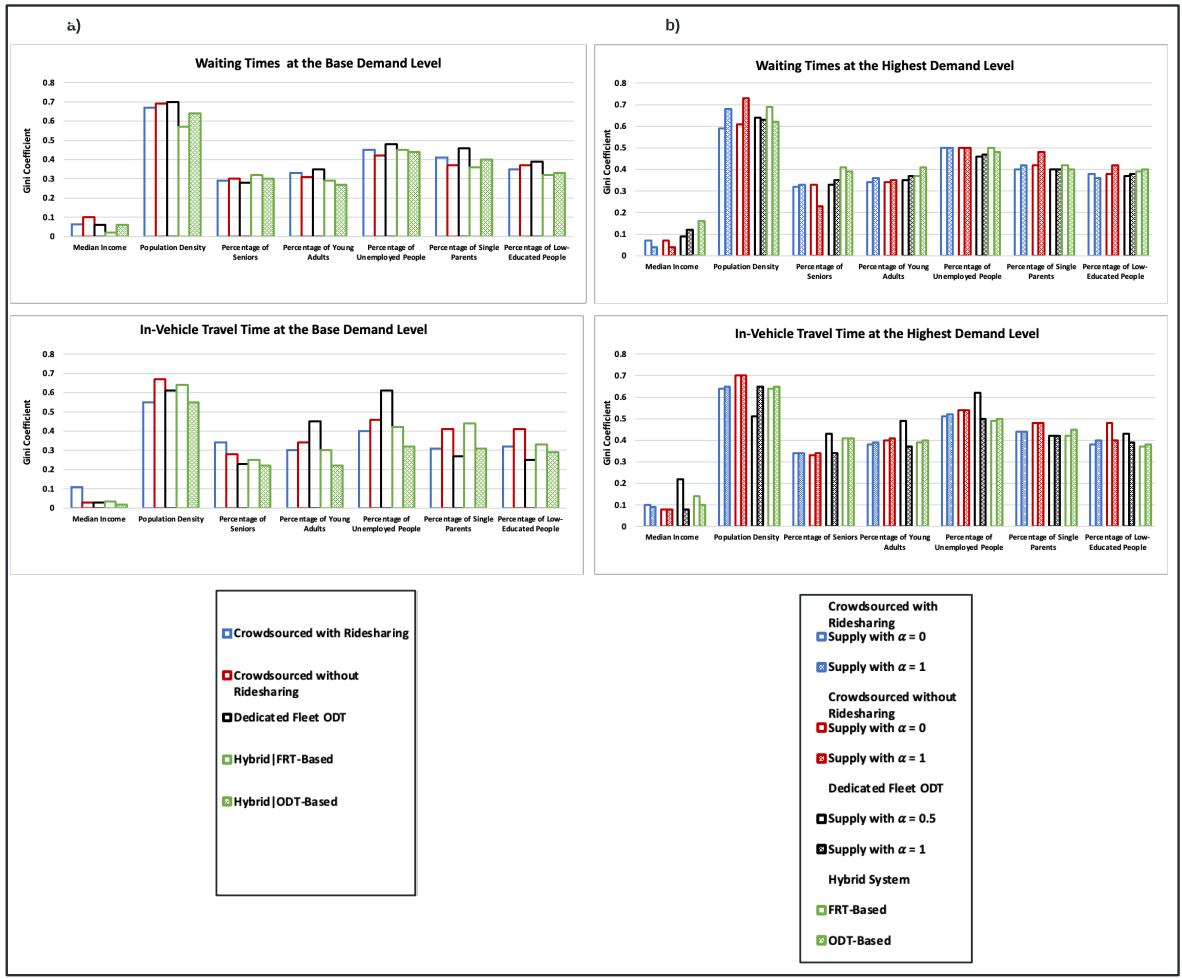

**Figure 7. Equity distribution in waiting times and in-vehicle travel times for simulated ODT configurations among disadvantaged communities.** a) Gini coefficients for the waiting and in-vehicle travel times at the base demand level. b) Gini coefficients for the waiting and in-vehicle travel times at the maximum demand level. It is worth mentioning that the analysis of the in-vehicle travel times are based on the dissemination area of the trip origin.

## Discussion and concluding remarks

The key elements of a sustainable transport system, according to previous literature, include being accessible, affordable, safe, environmentally responsible, cost-efficient, and socially equitable[16–19]. In this study, we provided a general framework for assessing the sustainability of ODT systems, making notable contributions to the existing literature in three key ways. Firstly, the framework evaluated the sustainability of ODT systems from multiple perspectives including overall efficiency, environmental footprint, and social equity and inclusion. While prior studies have examined some of these aspects individually for various transportation systems, this research is the first, to the best of the authors' knowledge, to collectively analyze all these dimensions within the context of ODT systems. Secondly, we proposed the use of the generalized cost value as a new performance metric for ODT systems. The new performance metric offers a holistic approach to evaluate the overall efficiency of ODT systems, identify demand switching points between different ODT configurations, and quantify uncertainties within the performance evaluation of ODT systems. Thirdly, we demonstrated the practical application of the proposed framework by applying it to a real transportation network. This empirical illustration offers transit agencies and decision-makers invaluable insights and recommendations concerning the development and implementation of sustainable ODT systems.

The proposed framework was applied to the Town of Innisfil, Ontario, where a crowdsourced ODT system has been implemented since 2017. A micro-simulation model was developed and calibrated on data from the ODT service. The transportation network of Innisfil was simulated under several ODT configurations, and the results were used to analyze their sustainability. The results indicated that the overall efficiency of the transit systems, measured by the generalized costs, depend on the supply settings and surge pricing level. In the case of adequate supply and no surge pricing, crowdsourced ODTs are the most cost-efficient transit system at all demand below 3.37 riders/$km^2$/day. However, when 20%, 40%, and 50% cost increments are applied to the operating costs of crowdsourced ODTs as surge pricing, the FRT-based hybrid systems becomes the most cost-efficient transit system when the demand is greater than 2.12, 1.30, and 1.18 riders/$km^2$/day, respectively.

In terms of the environmental footprint, the results revealed that the use of private vehicles is more environmentally sustainable than providing public transit service in Innisfil at all demand levels below 885 riders/day (3.37 riders/$km^2$/day). However, as the Town grows and the network becomes congested, higher capacity public transit vehicles will likely result in lower GHG emissions per passenger than private vehicles. Until that time comes, providing a public transit service remains imperative to serve disadvantaged communities, to help reduce congestion and parking problems, as well as to reduce residents' reliance on private vehicles in the future. These findings represent the most conservative scenario, where all vehicles are gasoline-powered. The environmental footprint of ODTs was also evaluated under five levels of electrification, from 20% to 100% with increments of 20% (Fig. 8). The electrification of the public transit fleet mitigates downstream GHG emissions on the road, but results in upstream GHGs associated with electricity production needed to charge vehicles[54]. Overall, having 20% of the fleet electric vehicles would result in a 19.6% reduction in total yearly GHG emissions from transit systems. The electrification of the entire fleet would save 98.1% of the total yearly GHG emissions. It is worth mentioning that the generation of electricity in Ontario is largely based on renewable energy sources (nuclear and hydroelectric) during off-peak hours[62]. Therefore, the implementation of optimized charging strategies that utilize the off-peak period can further reduce upstream GHG emissions from electric vehicles.

Our findings are transferable to most towns and suburban areas across North America, having similar land-use patterns, population densities, demographics, and transport-demand patterns. The transit agencies and policy makers can consider following key takeaways when planning new ODT systems.

- Crowdsourced ODTs:

    - With an adequate supply and no surge pricing, crowdsourced ODTs are the most efficient transit system for demand below 3.37 riders/$km^2$/day. High surge pricing could result in 50% increases in operating costs, thus the critical demand drops to 1.18 riders/$km^2$/day.

    - Incentive strategies for drivers are imperative for maintaining an adequate supply of vehicles as demand increases.

    - When there is a shortage of drivers, ridesharing can reduce waiting times for riders by up to 50% at high demand levels.

    - The existing crowdsourced ODT in Innisfil has low usage inequality rate across income, education, employment, young adulthood, seniors, and single parenthood. The social equity aspect of the service, however, can be further enhanced by providing more subsidies and discounts to low-income residents.

    - The existing crowdsourced ODT in Innisfil was able to withstand and recover quickly from the effects of COVID-19 pandemic (Supplementary Fig. 3).

    - Crowdsourced ODTs have the highest VKM per passenger and yearly GHG emissions, when compared with other designs.

- Dedicated fleet ODTs:

    - It has the highest capital and operating costs at all demand levels below 3.37 riders/$km^2$/day.

    - It is more cost-efficient than crowdsourced when the latter is in short supply and low surge pricing is applied (20% increment in the operating costs), and the demand is between 2.80 and 3.37 riders/$km^2$/day. When surge pricing is high, the dedicated fleet ODT becomes more cost-efficient for demand levels between 2.21 and 3.37 riders/$km^2$/day.

    - As demand grows, increasing the supply linearly with demand ($\alpha$=0.5) would be the most cost-efficient setting.

    - The dedicated fleet ODT with constant supply has the lowest VKM per passenger and yearly GHG emissions at all demand levels below 3.37 riders/$km^2$/day, when compared with other designs.

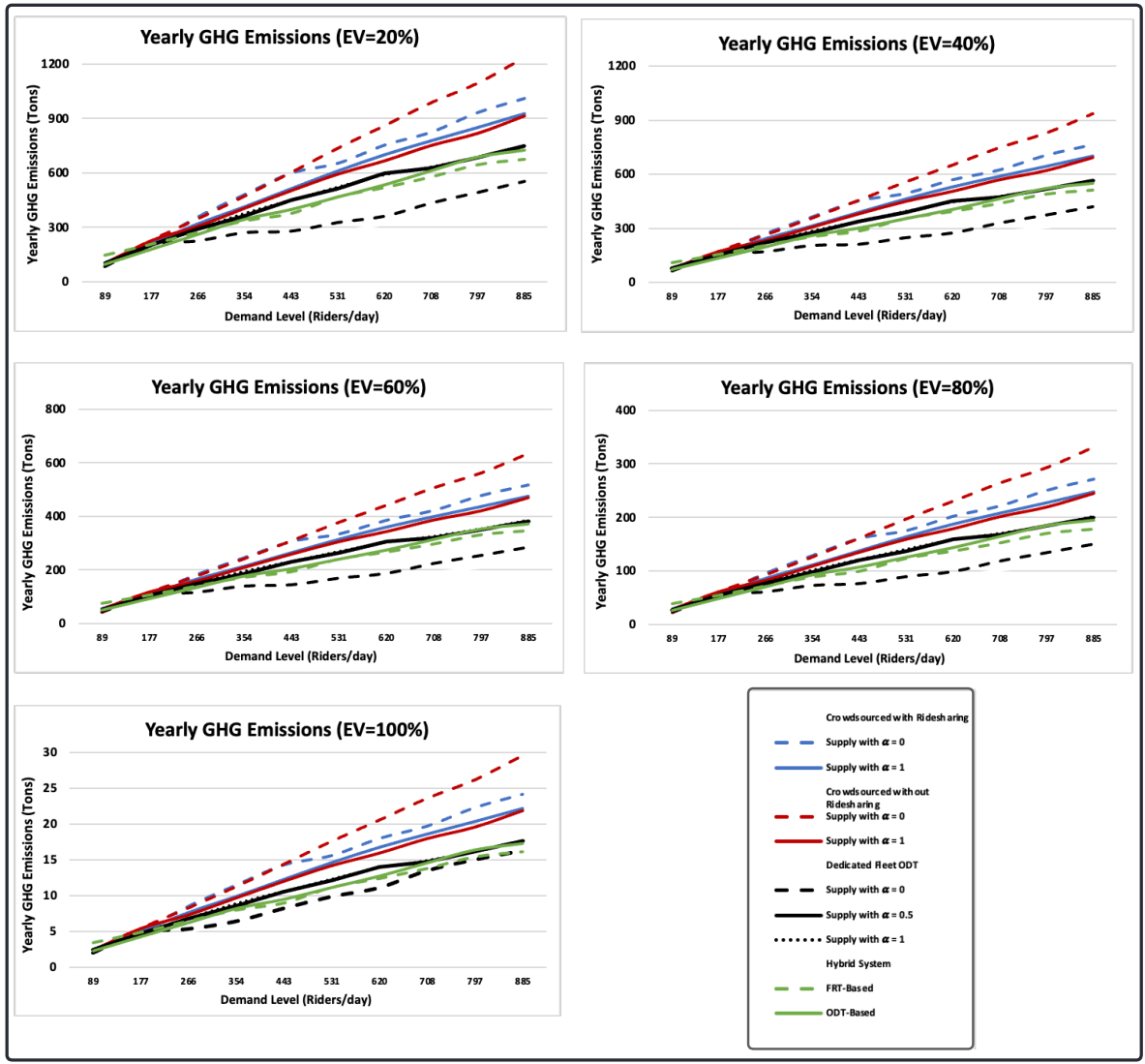

**Figure 8.** The environmental footprint of the simulated transit systems under five levels of electrification from 20% to 100% with 20% increments.

- Hybrid ODTs:
  - The FRT-based hybrid system is the most cost-efficient transit system to operate for demand greater than 3.37 riders/$km^2$/day. When high surge pricing is applied to crowdsourced ODTs, the critical demand value drops to 1.18 riders/$km^2$/day.
  - The FRT-based hybrid system has the second-lowest VKM per passenger and yearly GHG emissions for demand levels below 3.37 riders/$km^2$/day.
  - The performance of hybrid systems depends on the availability of an adequate supply of the crowdsourced ODT.
- General remarks:
  - The use of private vehicles is more environmentally sustainable than providing public transit service in Innisfil at all demand levels below 3.37 riders/$km^2$/day.
  - The electrification of the public transit fleet along with optimized charging strategies can reduce total yearly GHG emissions by more than 98%.

– Transit systems have similar equity distributions for waiting and in-vehicle travel times.

There are some limitations and potential future extensions to this study. We used online calculators and made some assumptions to calculate the costs and GHG emissions in this study. The results could be improved by obtaining more accurate data on surge prices, operating costs of FRTs and dedicated fleet ODTs, and emissions. Additionally, we only considered the environmental impacts of electrification. However, their impacts on fleet size, total costs, and critical demand densities were not captured, which could be of interest for future research. Furthermore, this study assumed that riders have homogeneous preferences across transit systems. In future studies, passengers' preferences and attitudes toward transit systems could be considered.

## Data Availability

The detailed operational data for the existing crowdsourced ODT in Innisfil (Uber Transit) is third-party data that we cannot upload directly. To access this dataset, researchers can contact Matthew Di Taranto, Senior Account Executive-Transit Partnerships, at (416-628-3234) or at (matthew.ditaranto@uber.com). The minimal dataset underlying the results described in our paper, including all the data needed to reproduce the figures, is available on a public repository at: https://github.com/LiTrans/Sustainability-Analysis-of-Shared-On-Demand-Public-Transit.git

## Acknowledgements

We are grateful to the Town of Innisfil and Uber Technologies Inc. for providing us access to the operational data of Innisfil Transit Service that was used in this study. We would like to thank Maryam Elbeshbishy for helping us in developing the simulation model. We are also grateful for the contribution of Paul Pentikainen from the Town of Innisfil, and Arjan van Andel from Uber Technologies Inc., for their inputs, suggestions, and review. This research was funded by the Town of Innisfil under grant number 1-51-48333 and Canada Research Program under grant number 2017-00038.

## Author contributions statement

The authors confirm their contribution to the paper as follows: **N.A.**: Conceptualization, Methodology, Data curation, Investigation, Formal analysis, Software, Visualization, Writing - original draft, and Writing - review & editing. **B.F**.: Conceptualization, Methodology, Investigation, Funding acquisition, Project administration, Resources, Supervision, and Writing - review & editing.

## Competing interests

The authors declare no competing interests.

## Additional information

Supplementary Information for: Sustainability Analysis of On-Demand Public Transit Systems.

# Supplementary Information for: Sustainability Analysis Framework for On-Demand Public Transit Systems

**Nael Alsaleh**[1] **and Bilal Farooq**[1,*]

[1]Laboratory of Innovations in Transportation (LiTrans), Department of Civil Engineering, Toronto Metropolitan University, Toronto, ON M5B 2K3, Canada
*bilal.farooq@torontomu.ca

## Supplementary Tables

| Results | | Waiting Time (min) | In-Vehicle Time (min) | Trip Length (km) |
|---|---|---|---|---|
| Actual | Average | 8.12 | 10.22 | 9.28 |
| | St.Dev | 5.17 | 6.65 | 4.11 |
| Simulated | Average | 8.72 | 11.39 | 9.76 |
| | St.Dev | 5.91 | 6.73 | 4.33 |
| Difference | Average | -0.60 | 1.17 | 0.48 |
| | St.Dev | 7.52 | 8.68 | 5.19 |
| | t-value | 1.04* | 1.79* | 1.23* |

*Not significant at 95% confidence level.

**Supplementary Table 1.** Statistical difference between actual and simulated performance. The comparison showed no statistical difference between the simulation results and the actual data, in terms of in-vehicle travel times, trip length, and waiting time values, at 95% confidence level.

# Supplementary Figures

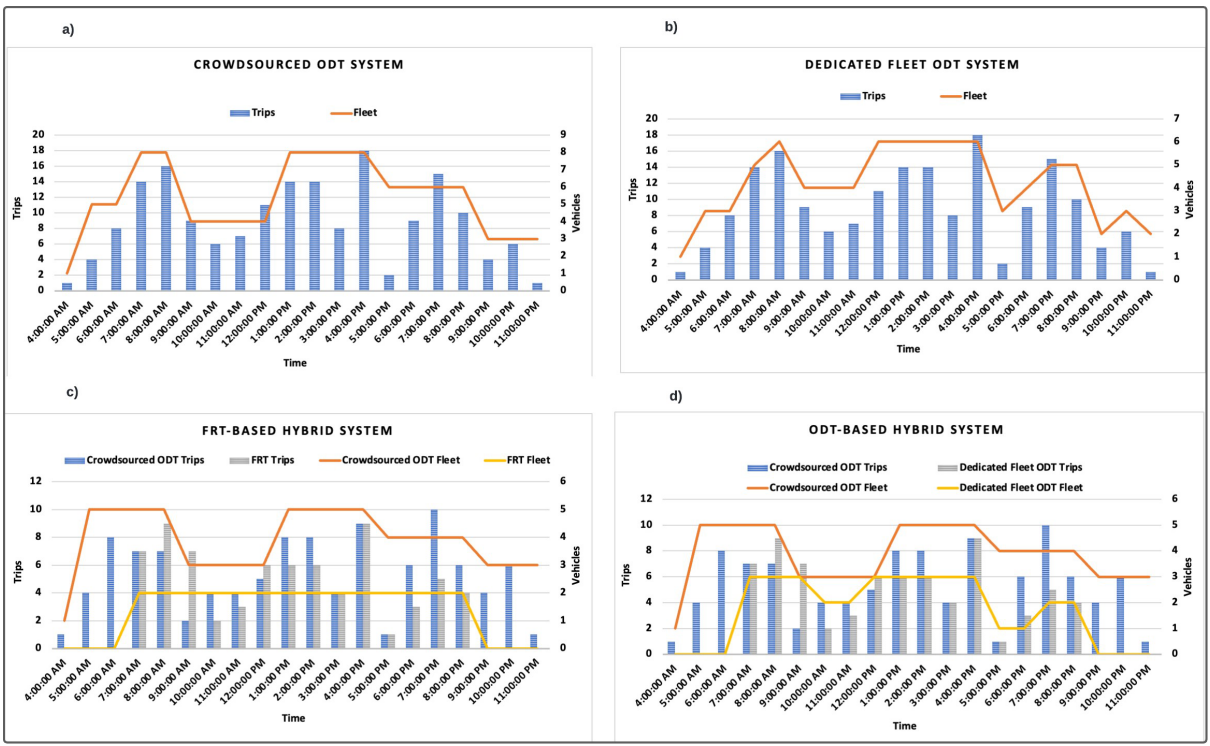

**Supplementary Figure 1.** The temporal distribution of demand and supply for a) the crowdsourced ODT system, b) dedicated fleet ODT system, c) the FRT-based hybrid system, and d) ODT-based hybrid system.

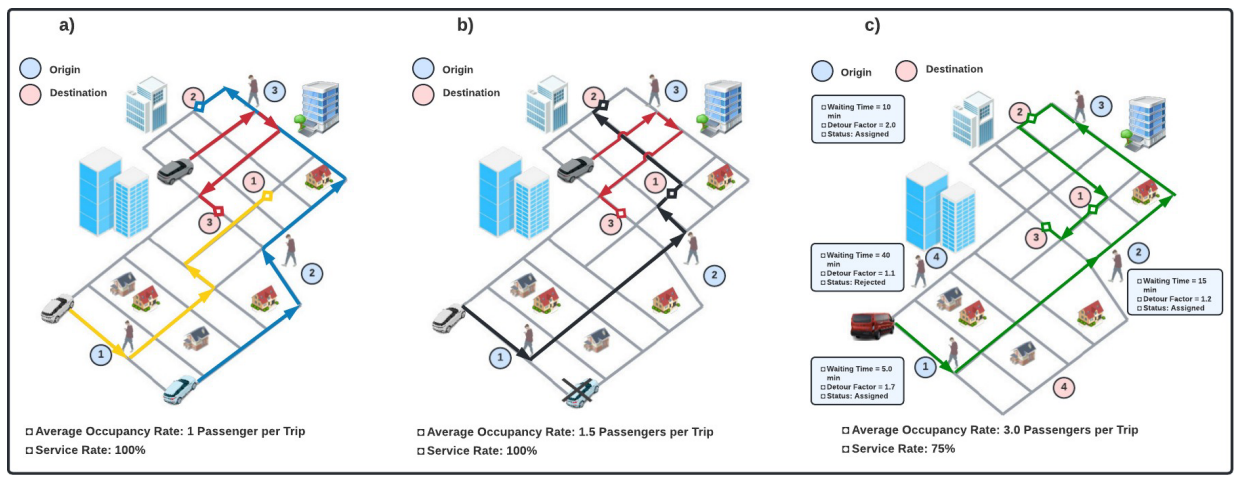

**Supplementary Figure 2.** Differences between the dispatching algorithms used in the crowdsourced ODT and the dedicated fleet ODT scenarios. a) crowdsourced ODT service with exclusive-rides. b) crowdsourced ODT service with shared-rides. c) Dedicated fleet ODT service.

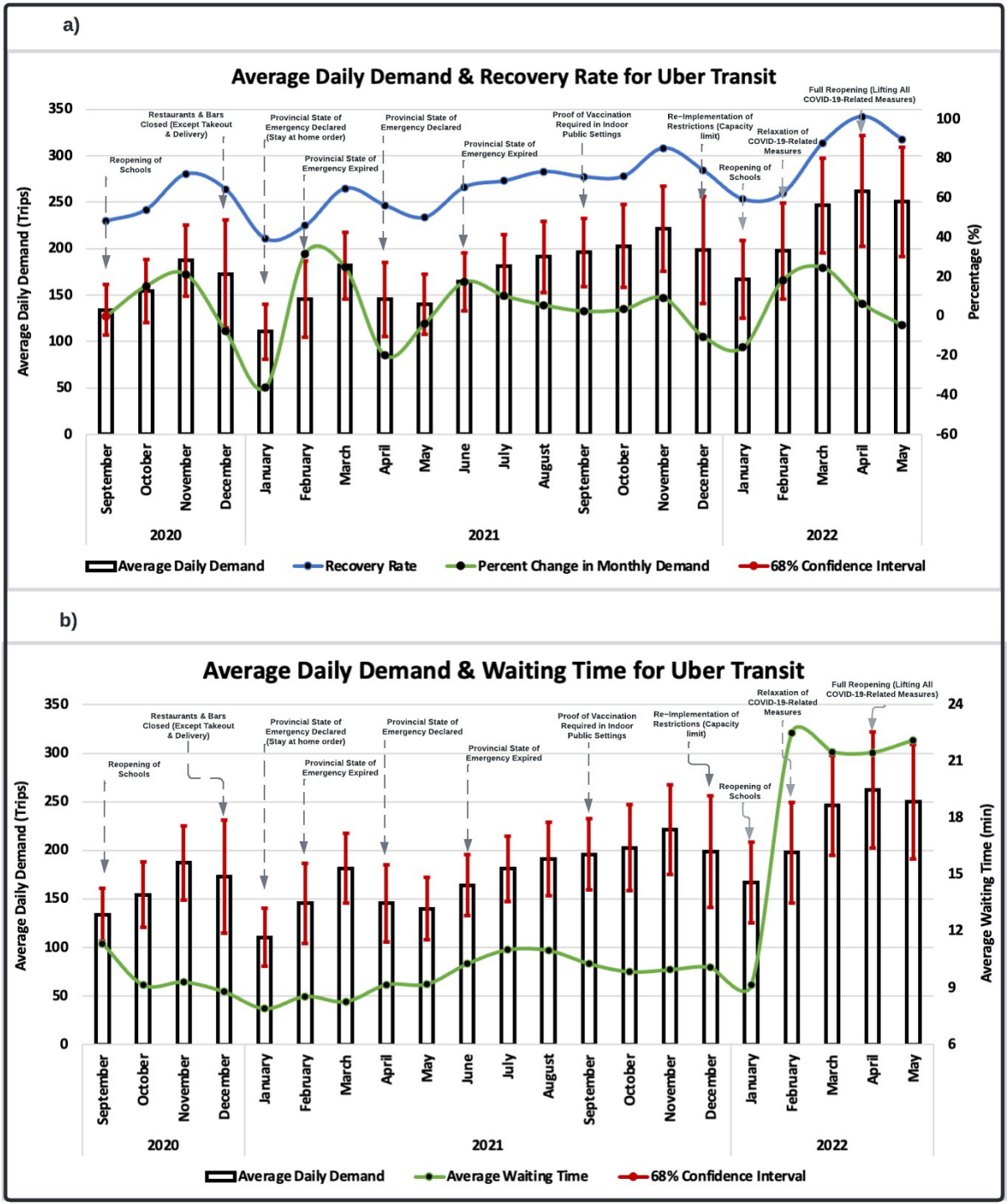

**Supplementary Figure 3.** The impact of the COVID-19 pandemic on the average daily demand and waiting time for the existing crowdsourced ODT in Innisfil. Demand recovery rates are calculated based on the 2019 demand levels. It is worth noting that the analysis was done on the data set from September 2020 in line with the effective start date of the data-sharing agreement between Innisfil and Uber Technologies Inc.

## Supplementary Notes

Supplementary Fig. 3 presents the average daily demand and waiting time for the service from September 2020 until May 2022. It is observed that the service experienced three cycles of a sharp decrease in demand followed by a gradual recovery. The highest reductions in demand were recorded in January and April 2021, when the province declared a stay-at-home order, at 58% and 41% of the 2019 levels, respectively. However, the service resumed its 2019 demand in April 2022 after all COVID-19 restrictions were lifted. On the other hand, the average daily waiting time values shown in Supplementary Fig. 3b illustrate the relationship between demand and supply. From September 2020 to January 2022, supply followed demand, rising and falling, resulting in consistent wait times between 8 and 11 minutes. In February 2022, the demand started to increase as a result of the relaxation of COVID-19 related restrictions. The increase in demand, however, was not accompanied by an adequate rise in supply, which led to a dramatic increase in the average waiting time. In March 2022, the Town of Innisfil launched an incentive program, which resulted in a slight decrease in waiting times in the following months.